\documentclass[3p,number]{elsarticle}
\pdfoutput=1
\usepackage[figuresright]{rotating}
\usepackage{afterpage,blindtext,graphicx,booktabs,tablefootnote,hyperref,mathtools,optidef}
\usepackage[T1]{fontenc}
\usepackage{textcomp}
\usepackage{berasans}
\usepackage{enumitem}
\usepackage[flushleft]{threeparttable}
\usepackage{multirow}
\usepackage[normalem]{ulem}
\useunder{\uline}{\ul}{}
\journal{Computer Communications}

\begin{document}

\begin{frontmatter}

\title{Benefits of Mobile End User Network Switching and Multihoming}

\cortext[cor1]{Corresponding author}
\author[aalto]{B. Finley\corref{cor1}}
\ead{benjamin.finley@aalto.fi}
\author[andes]{A. Basaure}
\ead{abasaure@miuandes.cl}
\address[aalto]{Department of Communications and Networking, Aalto University, Konemiehentie 2, Espoo, Finland}
\address[andes]{Facultad de Ingenier\'{i}a y Ciencias Aplicadas, Universidad de los Andes, Av. Monse\~{n}or \'{A}lvaro del Portillo 12455, Santiago, Chile}

\begin{abstract}
Mobile users have not been able to exploit spatio-temporal differences between individual mobile networks operators for a variety of reasons. End user network switching and multihoming are two promising mechanisms that could allow such exploitation. However these mechanisms have not been thoroughly explored at a general system level with QoE metrics. Therefore, in this work we analyze these mechanisms in a variety of diverse scenarios through a system level model based on an agent based modeling framework.

In terms of results, we find that in all scenarios end user network switching provides significant benefits in terms of both throughput and mean opinion score as the number of available networks increases. However, contrastingly, end user multihoming in most scenarios does not provide significant benefits over network switching given the same number of available networks. The major reason is inefficient radio resource allocation resulting from individual networks not taking the multihoming nature of end users into account. Though, in low user density situations this inefficiency is not a problem and multihoming does provide increased throughput though not increased mean opinion scores. Finally, scenarios that vary the fraction of users adopting multihoming suggests that both early and late adopters will have similar gains over users not adopting multihoming. Thus the adoption dynamics of multihoming appear favorable. Overall, the results support the applicability of end user network switching for improving mobile user experience and the applicability of end user multihoming in more limited situations.
\end{abstract}


\end{frontmatter}

%comment out line numbers for arxiv submission
%\linenumbers

\section{Introduction}\label{sec:introduction}
Mobile users increasingly expect an always-on high quality mobile connection regardless of their location. Mobile network operators (MNOs) have responded with technological advances such as LTE that have substantially increased connection throughput and reliability. However, so far, users have not been able to significantly exploit the temporal and spatial differences in quality between individual MNOs. This lack of exploitation partly results from the absence of widely available technical mechanisms, such as user driven national roaming or operator driven dynamic spectrum access, that allow such exploitation. 

The absence of operator driven mechanisms is primarily a result of regulatory uncertainty and the significant business and technical complexity of such schemes. Whereas the absence of user driven mechanisms is primarily a result of operator resistance as such mechanisms often require low switching costs\footnote{The currently high switching costs (enabled by the entrenched MNO model in many countries) result in users switching MNOs at a timescale of years. This switching frequency is orders of magnitude too slow to significantly exploit the spatio-temporal differences between mobile networks.} which potentially threaten operators current business models.

Given this operator resistance, mechanisms that do not require operator support are particularly interesting. In that vein, end user network switching is a mechanism that does not require operator support because the network switching is assumed to occur completely on the end user device. In addition, the related mechanism of end user multihoming (an end user transmitting over several networks simultaneously and thus aggregating capacity) does not require operator support given a higher layer multipath protocol such as MPTCP. The two mechanisms are fully defined in Section \ref{sec:definitions}. 

Given the potential of these mechanisms, a key driver for spurring adoption is understanding the scenarios in which these mechanisms will benefit users and whether the actual user benefit is substantial. However, prior work \cite{li2016,deng2016,li2013a,dandachi2016} on these mechanisms has primarily focused on low level technical implementations rather than higher level system analyses. Furthermore, these technical works have not applied user-centric QoE metrics such as mean opinion score (MOS) in their analyses. Therefore, in this work we examine these two mechanisms through a system level model that applies an agent based modeling approach. The model provides several end user performance metrics (including throughput and MOS) for a variety of diverse scenarios including both technical and market conditions such as layout of base stations, user densities, and rates of adoption of the analyzed mechanisms.

Since these mechanisms can be adopted without operator support, non-MNO ecosystem stakeholders such as consumers, handset vendors, mobile platform owners, and regulators should be particularity interested in understanding the benefits of such mechanisms.

We briefly describe the structure of the remainder of the paper. Section \ref{sec:definitions} gives brief definitions of end user network switching and multihoming, Section \ref{sec:switch_small_cell} details the related concepts of network switching costs and small cell operators, Section \ref{sec:related_lte_mechans} describes current related mechanisms in LTE, and Section \ref{sec:related_studies} details related work. Section \ref{sec:abm} introduces agent based models in general and Sections \ref{sec:overall_model}-\ref{sec:user_behavior_QoE_model} describes the specifics of the agent based model we use in this work including the network assumptions and agent behavior. Section \ref{sec:analysis} presents the results of the different scenarios. Section \ref{sec:network_assumptions} details the potential effects of the aforementioned network assumptions on our presented results. Finally Section \ref{sec:discussion} discusses the implications of results with a focus on regulators and Section \ref{sec:conclusions} gives brief conclusions.

\section{Background}\label{sec:background}
\subsection{Definitions}\label{sec:definitions}
Due to the lack of standardized terminology in this area we give brief definitions of the two end user mechanisms we analyze in this work.
\begin{itemize}
\item[] \textbf{End user network switching:} mechanism that allows a user to automatically and efficiently switch between mobile networks (with which the user has access through contracts) at a small timescale (scale on the order of seconds). The switching is performed entirely on the end user terminal and therefore no operator support is required.
\item[] \textbf{End user multihoming:} mechanism that allows a user to automatically, efficiently, and simultaneously use (transfer data over) several mobile networks (with which the user has access through contracts) thus aggregating network capacity. We assume that no information is shared between the networks (or between the BSs of the same network) about the multihoming nature of the users. This ensures that no operator support is required for end user multihoming.
\end{itemize}
Given these definitions we note that end user multihoming implies network switching to the extent that in end user multihoming the user simply selects the two best BSs (of all accessible to that user) to use simultaneously, whereas in end user network switching the user simply selects the best BS to use.

In terms of state of the art technical implementations of these mechanisms, \cite{li2016} describes a network switching implementation for off-the-shelf Google smartphones based on the Google Fi MVNO\footnote{Google Fi is a MVNO that aggregates three US MNOs (Sprint, T-Mobile, and U.S. Cellular) through dynamic switching of network SIM profiles.} that has an average switching time of 8.8s and a potential lower bound switching time of 1.5s. While \cite{deng2016} describes a multihoming implementation for off-the-shelf smartphones over LTE and Wifi with near optimal aggregation performance. We note though that the lack of off-the-shelf smartphones with two LTE stacks prohibits the current implementation of multihoming over two LTE networks.

\subsection{Switching costs and small cell operators}\label{sec:switch_small_cell}
Economically, switching costs are defined as one-time costs that a buyer faces when switching from one provider to another \cite{porter2008} and these costs constitute an entry barrier since they determine the monopoly power of incumbent firms. If switching costs are high, an entrant firm should attract new customers by subsidizing the customers switching costs. When switching costs are low, competition is more dynamic and a new firm can more easily enter the market.

The reduction of switching costs can incentivize in some cases the entrance of new types of operators. Specifically if switching costs are low enough and end users can efficiently switch from one network to another, the minimum efficient scale of an operator decreases. For example, an entrant operator can offer network access only in localized pockets given the assumption that users can easily and efficiently switch to a wider area operator outside those pockets. Such cases are especially interesting for new network deployments such as small cells, M2M and more generally IoT. This new type of operator is known as a small cell operator or micro-operator.

\begin{figure}[!ht]
    \includegraphics[width=16.4cm]{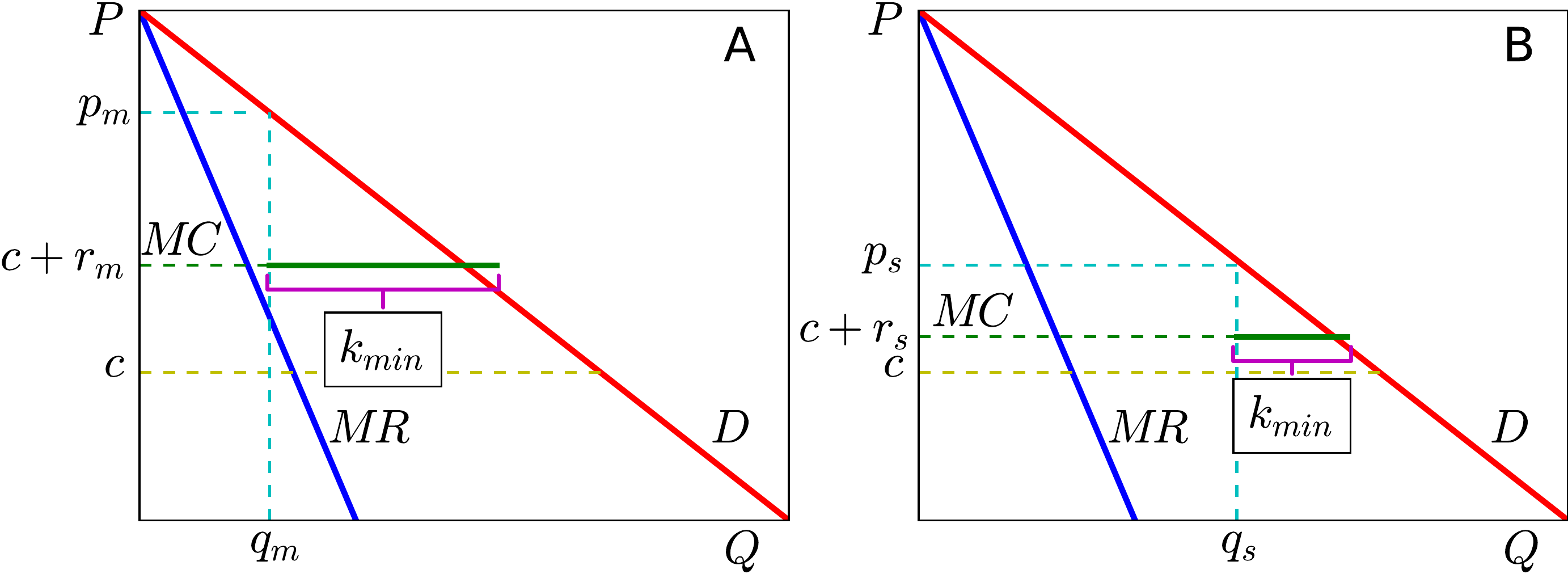}
    \caption{Demand $D$ (red), marginal revenue $MR$ (blue), and long run marginal cost $MC$ (green) curves under Spence-Dixit model illustrating: A) installed capacity $q_m$ and market price $p_m$ for the current situation given current $k_{min}$ (minimum efficient scale), $c$ is the marginal cost of production, $r_m$ and $c+r_m$ are respectively the unit cost of capacity and the long run marginal cost of production which considers both production and capacity costs given macro cells B) installed capacity $q_s$ and market price $p_s$ given smaller $k_{min}$, $c$ is the marginal cost of production, and $r_s$ and $c+r_s$ are respectively the unit capacity cost and long run marginal cost given small cells. In both panes the axes are price $P$ vs. quantity $Q$.}
\label{fig:spence_dixit} 
\end{figure}

From a competition perspective, according to Spence-Dixit capacity model \cite{spence1977,dixit1980}, the industry installed capacity in a market can act as an entry barrier to new firms. Such a situation is depicted in Figure \ref{fig:spence_dixit}A. Specifically, the industry installed capacity $q_m$ is an entry barrier in a market if $q_m$ is chosen such that the addition of the minimum efficient scale capacity $k_{min}$ (by an entrant) would not be profitable. This minimum efficient scale capacity is the minimum size at which an entrant operator is profitable (can recover its average costs) due to economies of scale. Graphically, in Figure \ref{fig:spence_dixit}A, this means that $k_{min}$ segment is too large to fit between current installed capacity $q_m$ and demand $D$; in other words, if a new operator enters with a capacity of $k_{min}$ the price will fall below the long run marginal costs.

If $k_{min}$ decreases, the industry can either increase capacity to again deter entrance given this new $k_{min}$ or allow entrance of new operators until this capacity is reached anyway. Industry will do whichever is more profitable, in either case, prices will decrease. Figure \ref{fig:spence_dixit}B illustrates the situation in which $k_{min}$ decreases thus pushing capacity to $q_s$ and price to $p_s$. Note that smaller deployments may have lower unit cost of capacity $r_s$ since they require less infrastructure.

\subsection{Comparison of end user mechanisms to related LTE methods}\label{sec:related_lte_mechans}
Several existing LTE mechanisms are conceptually related to end user multihoming, these mechanisms are briefly discussed.

Dual connectivity (DC) for LTE is a recent technical enhancement from 3GPP Release 12 (from 2012) that allows user equipment to receive data from either two BSs (specifically two eNodeBs of the same network) simultaneously or from the best BS of two connected BSs thus increasing user throughput and network efficiency \cite{zakrzewska2013}. The enhancement is often targeted at heterogeneous networks where the one of BS would be a macro BS and the other a small BS. Thus this enhancement is conceptually related to both end user multihoming and network switching, though several important differences remain. 

Primarily, DC is intended for cases where both BSs are from the same network and thus the enhancement often assumes information can be sent between the BSs through an inter-BS X2 interface. The X2 connection in DC can serve several purposes: sharing part of the actual user data (to be transmitted) with the secondary BS and bidirectionally sharing information about user throughput with each BS for scheduling purposes \cite{wang2016}. End user multihoming on the other hand never assumes a connection between BSs (of either different operators or of the same operator). In end user multihoming, the actual user data is assumed to be split between BSs through a higher layer protocol such as MPTCP, and scheduling is done independently by each BS. However we note that DC is a very flexible enhancement and in some variations of DC the data splitting is done at high layers and scheduling is independent on each BS \cite[Chapter~16]{dahlman2016}. 

Coordinated Multipoint (CoMP) is a technical enhancement related to DC but distinct in that in CoMP the multiple BSs transmitting to the user are often just distinct remote radio heads (RRH) that share the same baseband processing unit and thus the RRHs are assumed to have a virtually no latency connection between themselves \cite{irmer2011}. In other words, the CoMP case can be viewed as a special case of DC where the connection between the radio heads is essentially perfect (as compared to an X2 interface inter-BS connection). CoMP was developed before and is a precursor to DC.

\subsection{Related studies}\label{sec:related_studies}
Related work can be divided into studies with a similar focus on end user network switching and multihoming and studies with similar agent based modeling methodologies. 

\subsubsection{End user network switching and multihoming studies}
Several related studies have examined end user network switching and end user multihoming both from technical and economic viewpoints.

In the technical domain, iCellular \cite{li2016} is a novel end user network switching platform and algorithm. iCellular determines when to switch networks through a predictive machine learning (ML) approach (specifically when an alternative network is predicted to provide significantly better quality then a switch is initiated). Similarly, Delphi \cite{deng2016} addresses the problem of network switching with a general optimization framework and also allows for multihoming (over LTE and Wifi) via multipath TCP. The system was empirically tested in situations with both LTE and Wifi connectivity were available. Li et al. \cite{li2013a} also proposed a technique for distributing user flows in multihoming scenarios. The proposed technique accounts for trade-offs in different metrics including user QoE, cost, and energy.

Overall, most technically related studies have focused on the fundamental user problem of deciding how to divide traffic between the multiple network interfaces. This problem can be viewed as a complex multiuser multiobjective optimization problem wherein strategies will depend on the local and global information available to each user (i.e. technical network capabilities, current network loads, etc.) and the objectives of each user (maximize throughput, minimize energy usage, compromise of factors, etc.) in their current context. Ref. \cite{qadir2015} provides a survey of techniques. 

In contrast to these studies, we already assume that such traffic division can be done optimally (as we simply sum the capacities of available interfaces). Instead we include the lower level problem of network allocation of radio resources to devices. Towards this goal, Dandachi et al. \cite{dandachi2016} analyzed both user-centric traffic division and network-centric radio resource allocation schemes in the context of multihoming users. They illustrated that global resource allocation schemes that account for the multihoming nature of end users are more efficient that local schemes. Though they studied users multihoming over a single LTE and Wifi network rather than multiple LTE networks. 

In the economic domain, \cite{basaure2015b} compares the economic efficiency of end user multihoming against related schemes such national roaming and dynamic spectrum access and provides regulator guidance with respect to the three mechanisms. Also \cite{sonntag2015} performs a feasibility analysis of multipath protocols for IoT applications.

\subsubsection{Mobile network related agent based modeling studies}
Many mobile network studies have utilized agent based modeling. However the level of detail in both the agent and network components of these models varies greatly depending on the objective of the study. The most closely related studies to our work have been techno-economic studies with moderately complex agents yet relatively simple network components. For example \cite{basaure2015a} and \cite{basaure2015b} both employ such agent based models. 

Other more technically complex network models such as system level LTE simulators \cite{rupp2016} are in general too heavy\footnote{In other words, simulating three networks at 1ms granularity for many hours of usage with hundreds of users is impractical.} for our purpose and for the most part do not directly support, for example, users with connections to multiple mobile networks.

\section{Method}\label{sec:method}
\subsection{Agent based modeling}\label{sec:abm}
Agent based modeling is a general modeling methodology that employs a collection of typically rule based interacting agents in an defined environment. In contrast to strict equation based modeling, agents typically behave according to decision rules such that the model itself is not necessarily analytically tractable. Agent based models are especially advantageous in exploring complex systems since such models can often illustrate macro level emergent behavior through only simple micro level agent assumptions \cite{helbing2012} \cite{macal2010}. Thus since mobile networks and interacting autonomous users can be seen as a complex system, we argue that agent based modeling naturally fits well with the problem area. 

Several general agent based modeling platforms are open source and freely available including for example MASON\footnote{http://cs.gmu.edu/~eclab/projects/mason/} and FLAME.\footnote{http://flame.ac.uk/} In this paper, we utilize the Repast Simphony\footnote{http://repast.github.io} framework because the framework supports development in Java and the Java ecosystem provides a large variety of useful third party math and statistics libraries. In terms of implementation, we implement the model in about 2300 lines of Java code.

\subsection{Overall model parameters}\label{sec:overall_model}
Overall, the agent based model has several parameters that affect the general behavior of all sub-models with the most important being time. Specifically, time in the model progresses discretely with a one second granularity, and thus the entire model is updated and progresses (i.e. decisions are made) every one second. This time granularity provides an adequate level of network detail and realism while still allowing for the simulation of many agent hours in a reasonable amount of wall time. Furthermore, one second is an intuitive granularity for simulating user activities since user application sessions are often within a few orders of magnitude (a median smartphone session length is about 30 seconds \cite{finley2016a}).

As mentioned the model can be further broken down into sub-models including the network model, the user mobility model, and the user behavior and QoE model. These sub-models are described in detail in sections \ref{sec:network_model}-\ref{sec:user_behavior_QoE_model}.

\subsection{Network model}\label{sec:network_model}
The network model is essentially a high level system model of the radio access network. Thus we only model base stations and we consider backhaul and core network as out of scope.

The network model calculates the signal strength, bandwidth, and corresponding throughput available to a given agent based on the proximity of that agent to a BS(s) (of an operator with which the agent has a contract), the number of users using that BS(s), and the radio resource allocation scheme. We make several simplifying assumptions including that BSs operate at constant power all the time (thus we do not consider power optimization) and that fast fading smooths out at our one second time scale (and thus we can ignore it). We detail a variety of other network simulation parameters in Table \ref{tab:simulation_params}

\begin{table}[!ht]
\centering
\caption{Various network model parameters.}
\label{tab:simulation_params}
\begin{tabular}{|l|l|}
\hline
\textbf{Parameter} & \textbf{Value} \\ \hline
Outdoor transmission frequency (center) for macro cellular network & 2.1 GHz \\ \hline
Indoor transmission frequency (center) for small cellular network  & 3.5 GHz \\ \hline
Outdoor transmission carrier bandwidth for macro cellular network & 10 MHz \\ \hline
Indoor transmission carrier bandwidth for small cellular network  & 10 MHz \\ \hline
Outdoor macro cellular base station transmission power           & 37 dBm \\ \hline
Indoor small cellular base station transmission power        & 21 dBm \\ \hline
Log-Normal shadowing standard deviation                      & 6 dB \\ \hline
Outdoor BS height                                            & 30 m \\ \hline
Outdoor user equipment height                                & 1.5 m \\ \hline
\end{tabular}
\end{table}

Environmentally, the model defines two types of environments, the outdoor environment is served by macro cellular network and indoor environment is served by small cellular network, as depicted in Figure \ref{fig:scenario1_2}A. Specifically, in order to calculate the signal strength the network model utilizes a variety of empirical path loss models depending on whether the user or base station is indoors or outdoors. We detail this indoor/outdoor dichotomy and the corresponding path loss models in Table \ref{tab:path_loss_models}.

\begin{table}[!ht]
\centering
\begin{threeparttable}
\caption{Path loss models describing the downlink from base stations to the end user terminal.}
\label{tab:path_loss_models}
\begin{tabular}{|l|l|p{5.5cm}|p{5.5cm}|}
\hline
& & \multicolumn{2}{|c|}{\textbf{Base station}} \\ \hline
& & \multicolumn{1}{|c|}{\textbf{indoor}} & \multicolumn{1}{|c|}{\textbf{outdoor}} \\ \hline
\multirow{2}{*}{\textbf{User}} & \textbf{indoor}  & ITU indoor propagation model\tnote{a}  \cite{anderson2004}  & COST 231 Hata model for building penetration and non-light of sight\tnote{b} \\ \cline{2-4} 
& \textbf{outdoor} & COST 231 Hata model for building penetration and non-light of sight\tnote{c} & COST 231 Hata model for urban environment  \cite{cichon1999} \\ \hline
\end{tabular}
\begin{tablenotes}
      \small
      \item[a] The model assumes 1 internal wall for every 5m between the BS and user beyond 10m.
      \item[b] The model assumes 1 external wall and 1 floor.
      \item[c] The model assumes 1 external wall and 1 internal wall.
    \end{tablenotes}
\end{threeparttable}
\end{table}

All path loss models give a signal to interference plus noise ratio (SINR). This SINR along with the user bandwidth are then transformed to a throughput value via an approximate bounded and truncated Shannon function. The function (as dictated by appropriate adaptive coding and modulation schemes) is provided by 3GPP in Section A.2 of \cite{etsi2016}.

\subsubsection{Resource allocation scheme}
The potential radio resource allocation schemes of our simple network model fall in the category of frequency domain resource scheduling (FD-RS) since we allocate the bandwidth of each BS between the active users of that BS each second.\footnote{This essentially implies a time domain equal resource allocation since all active users are scheduled each second.} We select two simple and intuitive FD-RS schemes.

The base case allocation scheme is an equal resource (ER) scheme that simply divides a BSs bandwidth between all of the BSs users without regard to the channel quality (SINR) of users. This type of channel unaware scheme has been observed empirically in commercial networks \cite{hamad2015}. The alternative allocation scheme is essentially a throughput equalization (TE) scheme that divides a BSs bandwidth between all of a BSs users inversely proportionally to each users channel capacity (as calculated from SINR). This type of scheme represents a channel aware scheme. Both schemes attempt to impose a level of fairness and equality between users but with different metrics (bandwidth vs. throughput).

In our case, the ER scheme is also a locally\footnote{Locally means at the level of each individual BS} proportionally fair scheme. This can be shown relatively easily. Firstly, since each user is scheduled every second we only need to impose PF on each time slot independently. PF dictates that we maximize the sum of the logarithms of user's throughputs. This maximization problem for an individual time slot can be formalized as in Equation \ref{eq:prop_fairness} where $a$ is an individual user in the set of all users $A$, $t(\cdot)$ is the throughput function as detailed in Equation \ref{eq:throughput} that calculates throughput from the bandwidth allocated to the user $b_a$ and the SINR (in dB) of the user $s_a$, and $B$ is the total bandwidth of the cell (in our case 10 Mhz). Simple transformations (as shown in Equation \ref{eq:objective}) reduce the problem to the well known sum of logs optimization problem with a solution of $b_a=\frac{B}{|A|}$ (as used in ER).

\begin{maxi}|l|
  {b_a}{\sum\limits_{a \in A} {\log (t({b_a},{s_a}))}}
  {}{}
  \addConstraint{\sum\limits_{a \in A}{b_a}}{< B}
  \addConstraint{b_a}{> 0,\quad}{a \in A}
  \addConstraint{s_a}{> 0,\quad}{a \in A}
  \label{eq:prop_fairness}
\end{maxi}
 
\begin{equation}\label{eq:throughput}
t({b_a},{s_a}) = b_a*0.75*\log_2(1+10^{\frac{s_a}{10}})
\end{equation}

\begin{equation}\label{eq:objective}
\begin{aligned}
& \underset{b_a}{\text{max}} \sum\limits_{a \in A} {\log (t({b_a},{s_a}))} \\
& = \underset{b_a}{\text{max}} \sum\limits_{a \in A} {(\log(b_a) + \log(0.75*\log_2(1+10^{\frac{s_a}{10}})))} \\
& \equiv \underset{b_a}{\text{max}} \sum\limits_{a \in A} {\log (b_a)}
\end{aligned}
\end{equation}

In the context of multihoming, since we assume that users use the full amount of bandwidth provided, the splitting of traffic between the network interfaces (which we assume occurs on a higher layer) does not need to be considered. In practice this would be a user-centric allocation with the splitting depending on the specific higher layer protocol (such as MPTCP).

\subsubsection{BS selection scheme}
In terms of BS selection, every second each user selects the BS (or BSs) with the best estimated throughput given the BS selections of all other users. Thus the first user to make the selection in a given second has no information about others user's selection, while the last user has complete information (since all other users have already selected). The order of this selection process is randomized every second to remove any unfairness. This ordered selection procedure is analogous to a temporal ordering of requests from users in a given second. The estimated throughput takes into account both the BSs already selected by other users and the radio resource allocation scheme and therefore users have essentially perfect information.

In the context of BS selection research, our scheme represents a distributed load-aware user-driven approach as classified by \cite{wang2015}. Load-aware schemes are not currently a part of LTE, however such schemes are likely to be included in future 3GPP (i.e. 5G) releases \cite{wang2015,ramazanali2016}. In practice we note that BS load information could be broadcast by each BS in the broadcast channel \cite{wang2015} or potentially estimated by end user devices (without BS assistance) through techniques similar to \cite[Section 4.2]{xie2017} (which uses RSRQ-based estimation).

\subsection{User mobility model}\label{sec:mobility_model}
The user movement model uses a levy walk \cite{rhee2011} with truncated power law flight lengths (and five second pause times between each movement). The movement speed of users is 2 m/s (7.2 kph) outdoors (brisk walking) and 0.2 m/s (0.72 kph) indoors to account for the typically faster movement of users while outdoors.

\subsection{User behavior and QoE model}\label{sec:user_behavior_QoE_model}
\begin{figure}[!ht]
  \includegraphics[width=16.4cm]{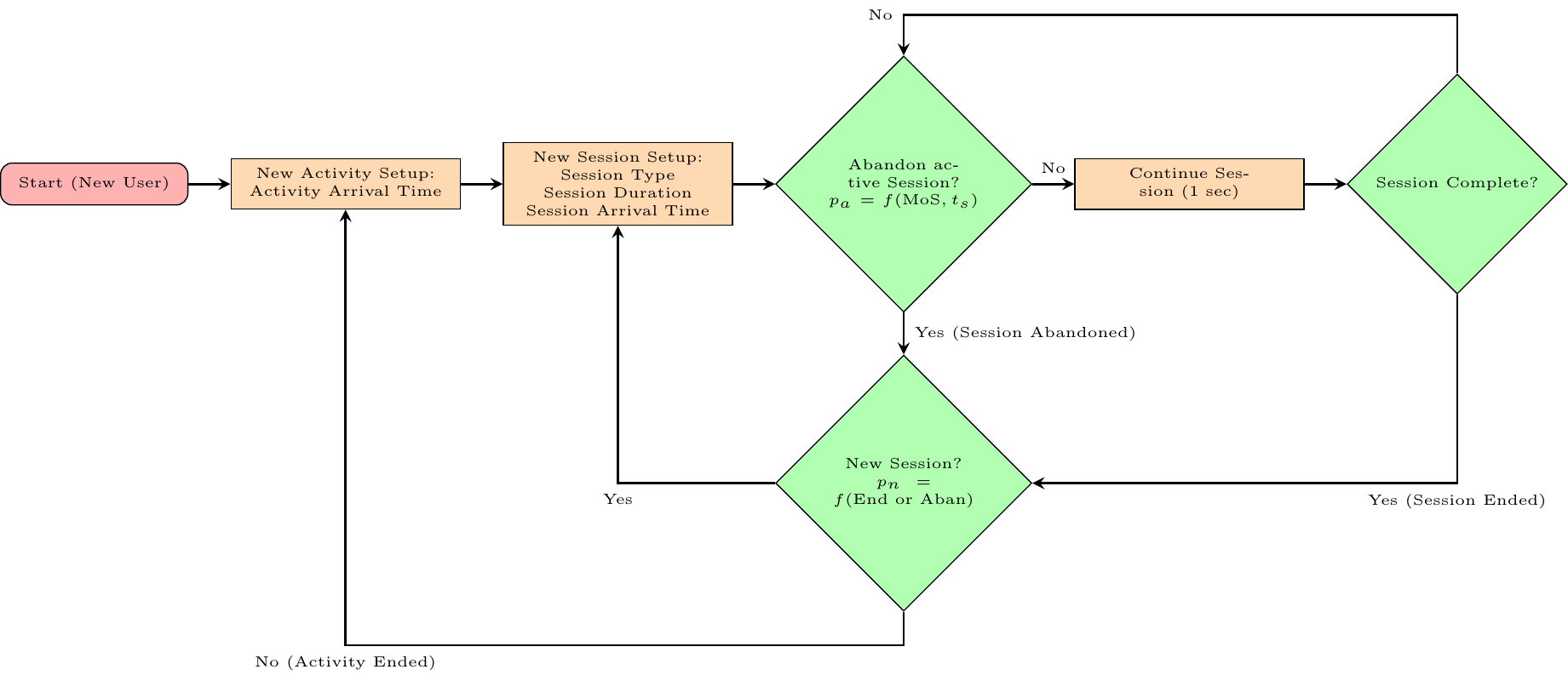}
  \caption{User behavior and QoE model flow chart. Note the probability of session abandonment $p_{a}$ is a function of the MoS at the given time $t_{s}$, while the probability of a new session $p_{n}$ is a function of whether the previous session ended normally (End) or was abandoned early (Aban).}
  \label{fig:user_model_flow_chart}
\end{figure}

The user behavior and QoE model dictates the user's mobile application usage behavior and the resulting user's evaluation of quality of the session. The model was adapted from \cite{tsompanidis2014} with modifications for using MOS as a user QoE metric (in addition to throughput). In the model, the behavior of each user is modeled as a Poisson burst process with each burst consisting of a single activity. Each activity therein consists of one or more application sessions. The model is illustrated as a flow chart in Figure \ref{fig:user_model_flow_chart} and the probability distributions and functions used in the model are summarized in Table \ref{tab:user_behavior_model} in the Appendix. We note that in order to introduce user heterogeneity many of the distribution parameters specified for each user are themselves samples from normal or truncated normal distributions. We further describe the dynamics of activities and sessions in the model in sections \ref{sec:session_model}-\ref{sec:activity_model}.

\subsubsection{Session}\label{sec:session_model}
Each session in the model is assigned a specific application type. The types we consider are general web browsing, messaging, social networking, video, and maps.\footnote{We consider these application types because they are all very popular and prior research provides QoE results for these types \cite{casas2016a}.}

The duration distributions of session types are derived empirically from actual user session data. Specifically, the distributions are estimated empirically via the Powerlaw package \cite{alstott2014} on the active smartphone dataset described in \cite{finley2016b}. In each case the distribution is selected based on Akaike weights.

\begin{figure}[!ht]
    \centering
    \includegraphics[width=16.4cm]{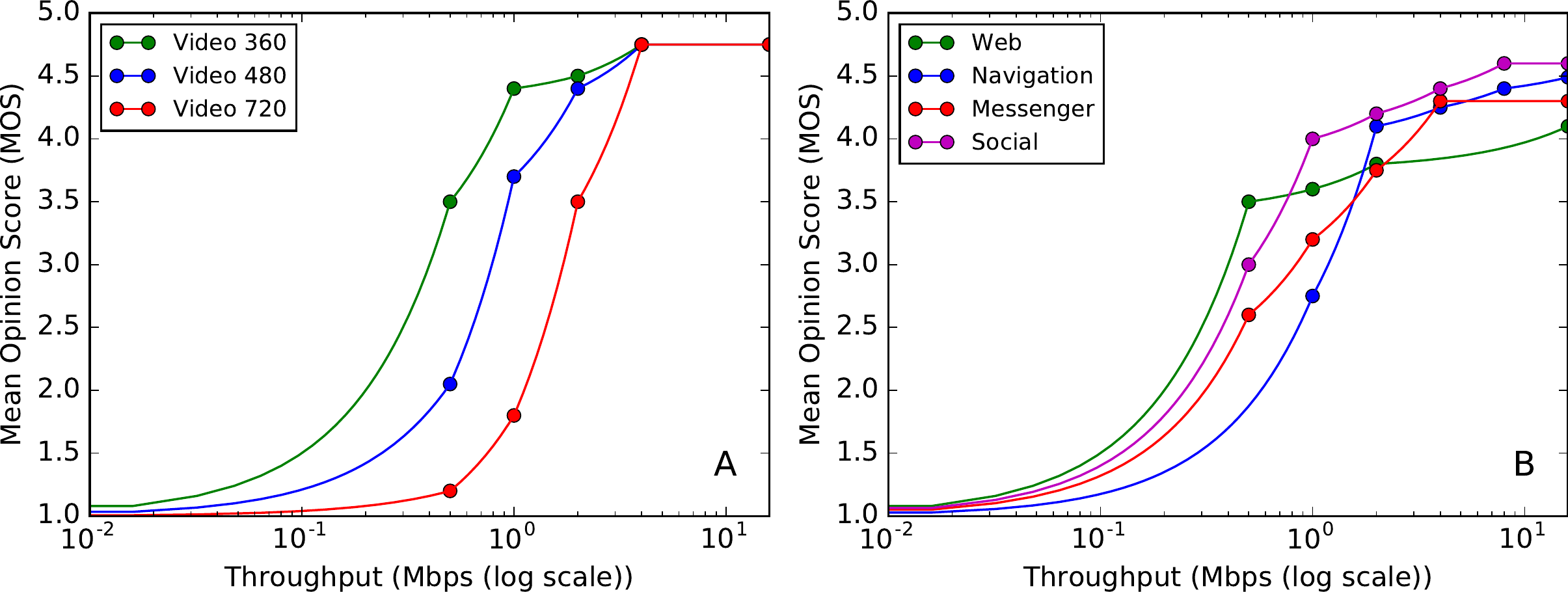}
    \caption{Throughput-to-MOS mapping functions as linear spline interpolations for different applications: A) video with resolutions \textless 720p and B) all non-video applications.}
    \label{fig:throughput_to_mos_mappings} 
\end{figure}

The required throughput for sessions of a given type to provide certain quality to the user (as measured by MOS) is defined by specific throughput-to-MOS mapping functions. The throughput-to-MOS mapping functions are derived from experimental data from \cite{casas2016a}. Specifically, for all applications except video with resolution \textgreater 720p, throughput to MOS mapping points are extracted from the results and splines with linear basis functions are fit given the mapping points as knots. We illustrate these fitted splines for the different session types in Figure \ref{fig:throughput_to_mos_mappings}. For video with resolutions of 1080, 2K, and 4K experimental results are not available and instead splines are fit based on a heuristic given in \cite{casas2016b} (specifically the $\beta$ rule). The $\beta$ rule requires estimates of the throughput needed for smooth (non-buffering) video playback for these resolutions. We utilize estimates of 8, 15, and 20 Mbps for 1080, 2000, and 4000 resolutions respectively.

Delay is also a factor for QoE (and therefore MOS) in certain applications \cite{casas2016a}. However we don't consider delay in our analysis because this would require a significantly more complex network sub-model and experimental data is not available for delay-to-MOS or (delay, throughput)-to-MOS combinations for most user applications we consider. Additionally, the experimental data that is available suggests that delay has a somewhat smaller effect than throughput \cite{casas2016a}. Therefore, we leave the analysis of delay for future research.

Additionally, sessions can be abandoned early if the quality of the session is too low. For every one second of a session where the cumulative mean MOS of the session is below a threshold value of 2.5 the user has a probability of abandoning the session. This abandonment probability is inversely proportional to MOS with a maximum of 5\% (at $\text{MOS}=1.0$) and minimum of 0\% (at $\text{MOS}=2.5$). This abandonment cannot occur during the initial grace (or start-up) period of the session which we define as the first 5 s. If a session is abandoned early then the probability of an additional session (in the same activity) is affected as described in section \ref{sec:activity_model}.

Finally, the probability of any given session being assigned a specific type depends on a weighted enumerated distribution over the types.\footnote{In statistical terms a discrete probability distribution with a finite sample space} The weights (and thus probability of selecting each type) are detailed in Table \ref{tab:user_behavior_model}. Additionally, for video sessions the video resolution is also selected from a weighted enumerated distribution with probability values for specific resolutions as detailed in Table \ref{tab:video_resolutions} in the Appendix. The mix of video resolutions is weighted towards a somewhat futuristic scenario with a significant fraction of video at resolutions of $\geq 1080$.

\subsubsection{Activity}\label{sec:activity_model}
As mentioned each activity contains at least one session (the initial session), however the probability of an additional session (in the same activity) following any given session is dictated by a binomial distribution with a mean over all users of 50\% probability given the session was completed or a 20\% probability given the session was abandoned early. Furthermore, if an additional session is required the inter-session time between the completed session and the additional session is dictated by a Pareto distribution. Finally, if the last session of an activity was abandon early then the activity itself is considered a failure and the time to next activity distribution is affected (see Table \ref{tab:user_behavior_model} in the Appendix).

\subsubsection{Overall user MOS}
The overall experience of the user is derived as the mean MOS over all of a user's sessions in the simulation. However, the early abandonment feature of our model causes an issue because sessions with poor MOS that are abandoned early are less costly than sessions with poor MOS that are not abandoned early. This is not likely the case in reality and thus a penalty must be used to account for the user frustration of early abandonment. Thus we calculate the mean MOS value as if early abandoned sessions were the full duration but with a MOS score of 1.0 for the abandoned time of the session.

\section{Scenarios}\label{sec:analysis}
\subsection{Simulation procedure}
\begin{figure}[!ht]
  \includegraphics[width=16.4cm]{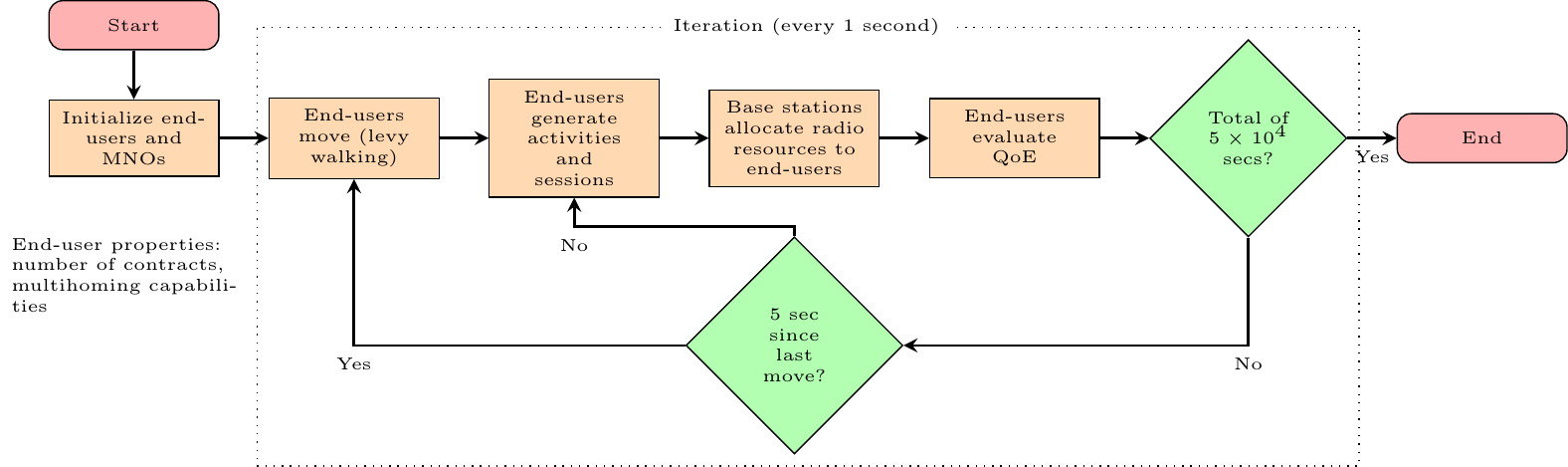}
  \caption{High level simulation flow chart.}
  \label{fig:complete_model_flow_chart}
\end{figure}

For each scenario we simulate (via the agent based model) 50,000 s (13.88 h) of user action as we find this duration allows a good convergence of the simulation results. Figure \ref{fig:complete_model_flow_chart} illustrates a high level flow diagram of this simulation process. We also repeat each simulation 10 times with different random number seeds to obtain 10 result sets. These result sets are then used to obtain normal 90\% confidence intervals for all the results. However, we note in some cases the confidence intervals are too small to be included in certain figures, in these cases we only display the mean of the result sets. Each scenario simulation was performed on a standard desktop computer with Intel Core i7 (2.5 Ghz) and 16 GB RAM and took between 2 and 45 min wall time depending primarily on the number of users in the scenario (between 25 and 425).

\subsection{Baseline scenario}
The baseline scenario (also referred to as scenario 1) is the default scenario upon which we vary certain parameters to obtain all other scenarios. The baseline scenario includes 3 different operators and 300 users equally split between the operators. Spatially, as illustrated in Figure \ref{fig:scenario1_2}A, the scenario dictates a 2.56 $km^2$ area (in white) with 6 outdoor BSs and a centrally located 3600 $m^2$ indoor area (in dark grey) with 6 indoor BSs. Users move within a 0.09 $km^2$ area (in light grey) that encompasses and includes the central indoor area. Operator base stations are positioned such that they are symmetrically located on opposite sides of the areas. Additionally, in this baseline scenario, all users have only a single operator (and thus do not perform network switching or multihoming). Therefore this represents the current dominant situation in terms of network switching and multihoming. Finally, the radio resource allocation scheme is the proportionally fair ER scheme.

The baseline scenario represents an urban point-of-interest such as a train station, concert hall, or other crowded public place that would likely include outdoor and indoor base stations. We focus on urban scenarios because we assume the benefits of network switching and multihoming in rural areas, though interesting, are likely more straightforward and dominated by coverage differences rather than an interplay of factors.

\setlength{\unitlength}{1.0cm}

\begin{figure}[!ht]
\begin{minipage}{0.50\textwidth}
        \centering
        %\fbox{
        \begin{picture}(7.7,7.7)
        \put(0,0){\includegraphics[width=7.7cm]{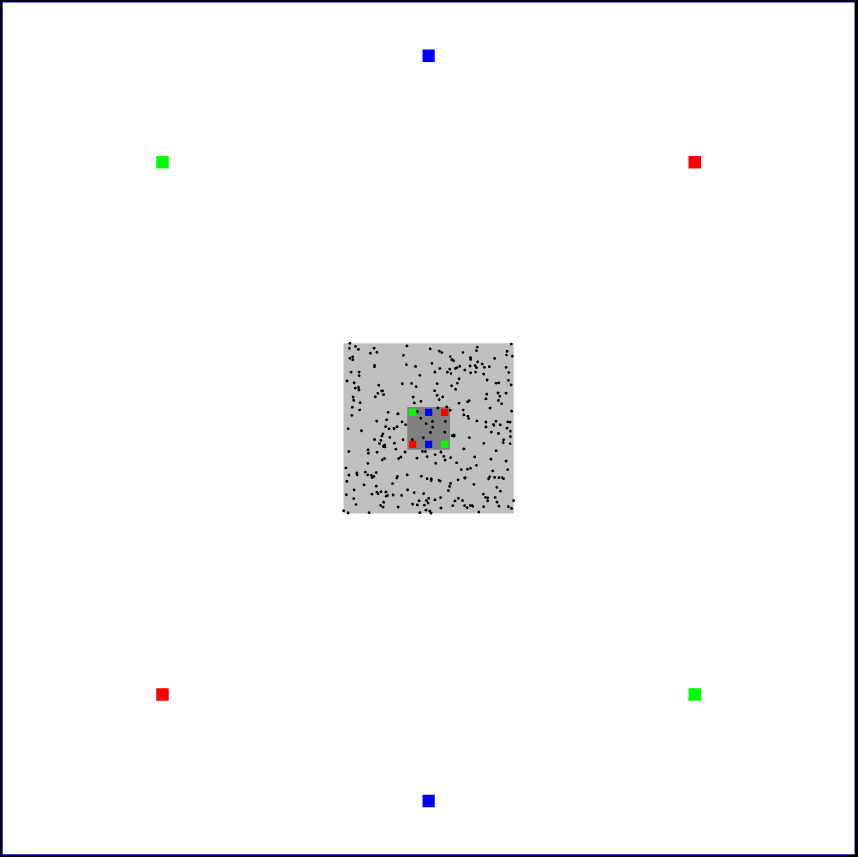}}
        \put(0.4,7.0){{\fontfamily{fvs}\selectfont A}}
        \end{picture}
        %}
    \end{minipage}\hfill
    \begin{minipage}{0.50\textwidth}
        \centering
        \begin{picture}(7.7,7.7)
        \put(0,0){\includegraphics[width=7.7cm]{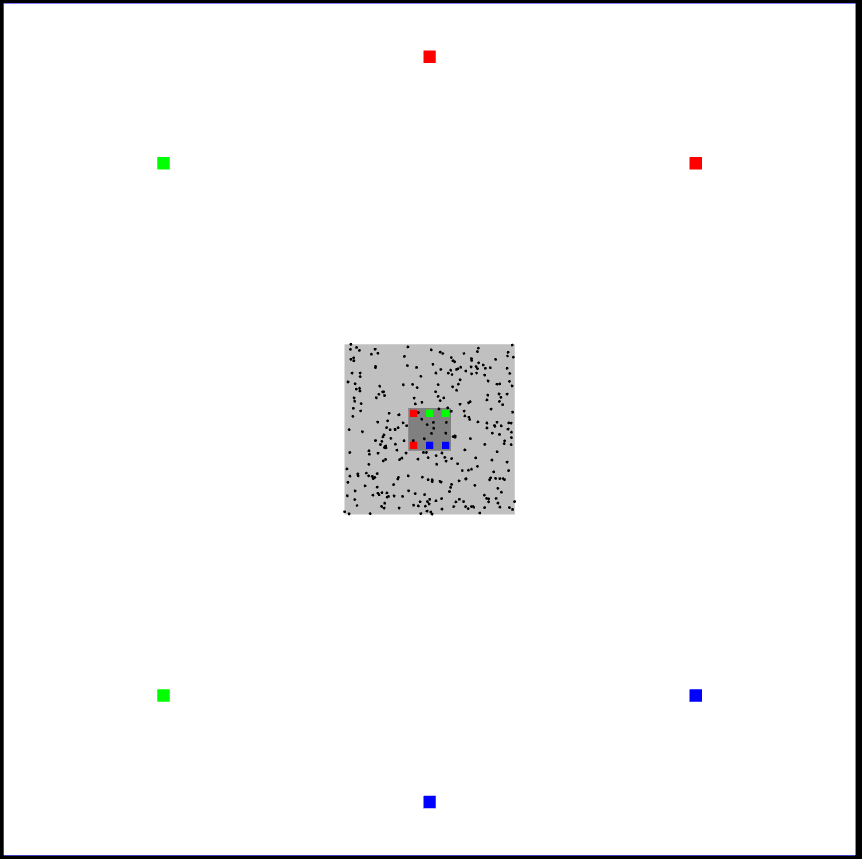}}
        \put(0.4,7.0){{\fontfamily{fvs}\selectfont B}}
        \end{picture}
    \end{minipage}
    \caption{Network layout scenario illustrations with colored squares representing base stations of different operators, light grey area representing the bounded area that users move within, dark grey area representing indoor space (that users also move in), and black dots representing individual users. A) scenario 1 (baseline) and B) scenario 2 (asymmetric layout).}
    \label{fig:scenario1_2} 
\end{figure}

\begin{figure}[!ht]
\begin{minipage}{0.50\textwidth}
        \centering
        \begin{picture}(7.7,7.7)
        \put(0,0){\includegraphics[width=7.7cm]{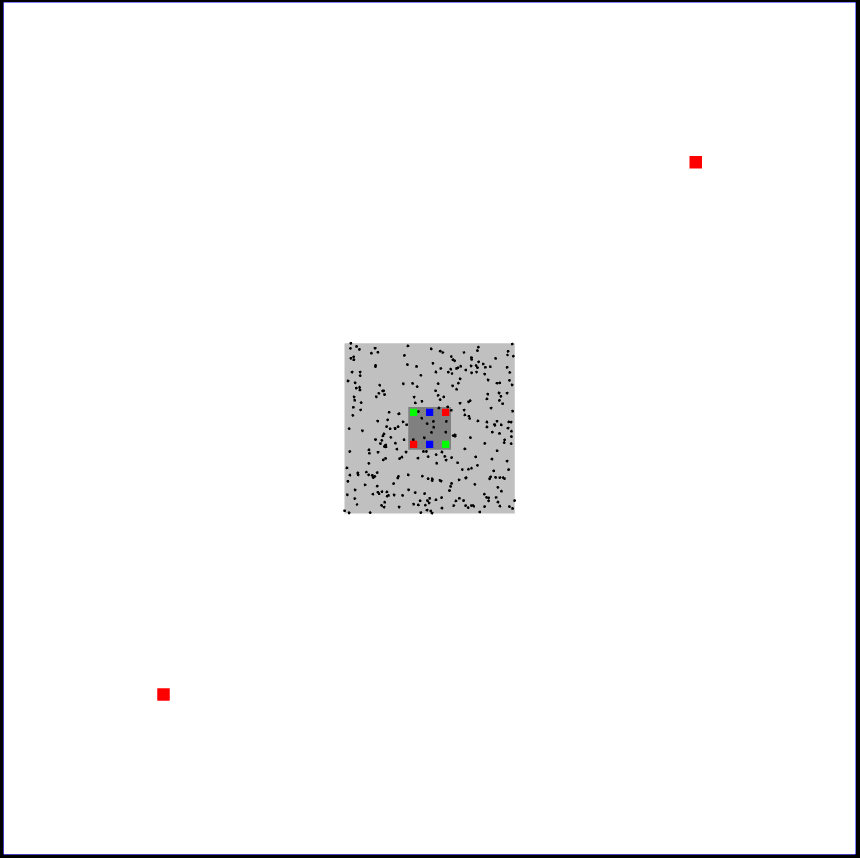}}
        \put(0.4,7.0){{\fontfamily{fvs}\selectfont A}}
        \end{picture}
    \end{minipage}\hfill
    \begin{minipage}{0.50\textwidth}
        \centering
        \begin{picture}(7.7,7.7)
        \put(0,0){\includegraphics[width=7.7cm]{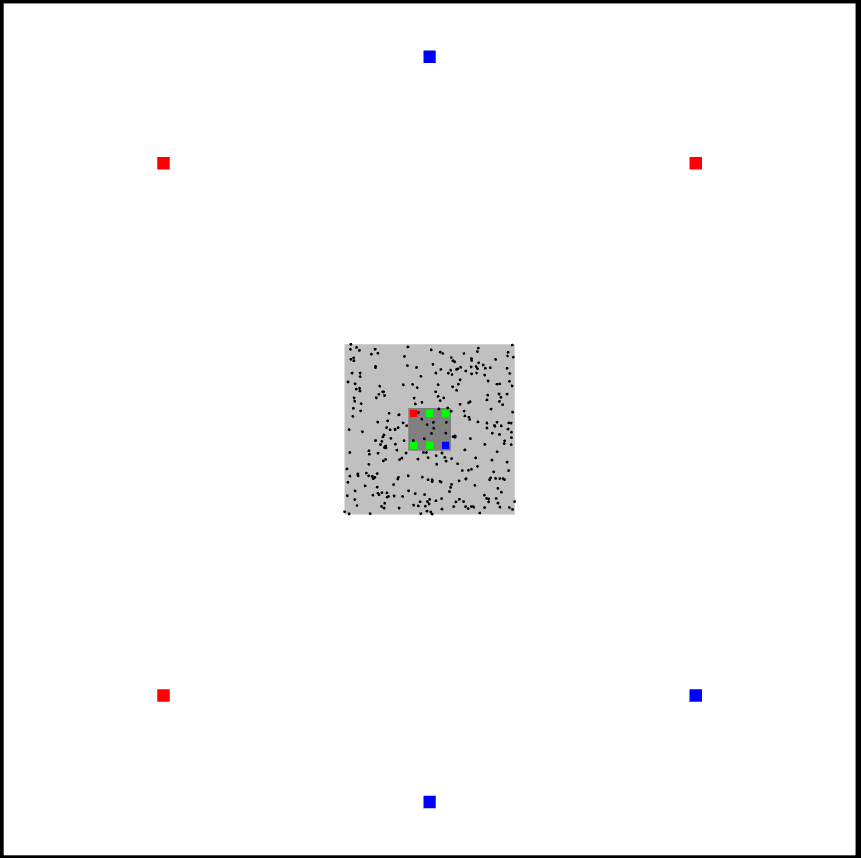}}
        \put(0.4,7.0){{\fontfamily{fvs}\selectfont B}}
        \end{picture}
    \end{minipage}
    \caption{Network layout scenarios illustrations with colored squares representing base stations of different operators, light grey area representing the bounded area that users move within, dark grey area representing indoor space (that users also move in), and black dots representing individual users. A) scenario 3 (co-located outdoor BSs - note that outdoor BS are overlapped such that each operator has a BS in each of the two outdoor locations) and B) scenario 4 (indoor only green operator, refer to web version for color figure).}
    \label{fig:scenario3_4} 
\end{figure}

\subsection{Parameter variations}
\begin{itemize}
\item[] \textbf{Network layout:} We can move the base station positions of operators or change the number of base stations per operator to acquire different network layouts. For example, we can create layouts with co-located BSs or non-symmetric BSs. Four distinct network layout scenarios are defined (scenarios 1-4). These layouts are illustrated in Figures \ref{fig:scenario1_2} and \ref{fig:scenario3_4}.
\item[] \textbf{Radio resource allocation scheme:} We can utilize the channel aware TE radio resource allocation scheme instead of the channel unaware ER scheme.
\item[] \textbf{Number of end users:} We can vary the number of end users thus creating different user density levels.
\item[] \textbf{Network switching and multihoming user fractions:} We can vary the number of contracts per user (thus effecting network switching) and the ability for each user to use multihoming. Given these two dimensions, and the restriction that multihoming requires 2 contracts, we have 5 potential user types: 1 contract (implies no multihoming), 2 contracts (no multihoming), 2 contracts (multihoming), 3 contracts (no multihoming), and 3 contracts (multihoming). Thus we can vary the percentages of these user types subject to the requirement that they sum to 100\%\footnote{We always split all user types equally among the operators.}. We abbreviate these user types as 1C, 2C-NMH, 2C-MH, 3C-NMH, 3C-MH respectively. 
\end{itemize}

\subsection{Results}
\begin{figure}[!ht]
    \includegraphics[width=16.4cm]{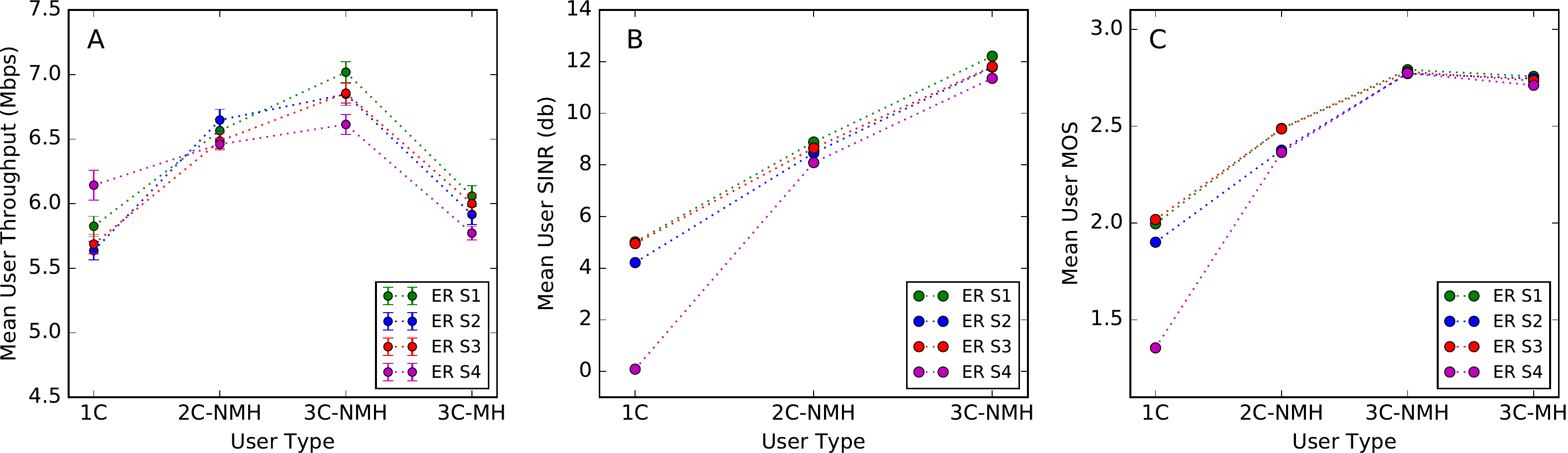}
    \caption{Performance values for different scenarios and user types (all users are a single type) given ER radio resource allocation scheme: A) mean user throughput B) mean user SINR C) mean user MOS.}
    \label{fig:er_all_scenarios} 
\end{figure}

\begin{figure}[!ht]
    \includegraphics[width=16.4cm]{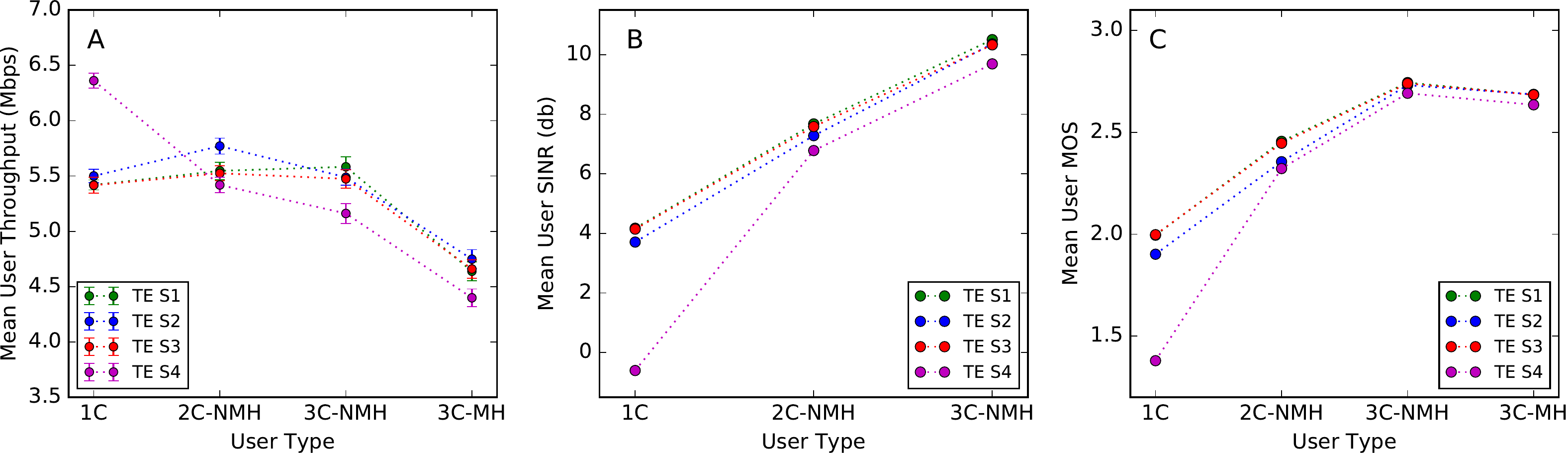}
    \caption{Performance values for different scenarios and user types (all users of a single type) given TE radio resource allocation scheme: A) mean user throughput B) mean user SINR C) mean user MOS.}
    \label{fig:te_all_scenarios} 
\end{figure}

First, to assess the overall benefits of complete (100\%) adoption of network switching or multihoming we consider situations in which all users are of a single user type (i.e. 2C-NMH or 3C-MH). For brevity we exclude the 2C-MH type from analysis and focus on 3C-MH type as the results are similar. Figures \ref{fig:er_all_scenarios} and \ref{fig:te_all_scenarios} illustrate the mean performance values\footnote{We note that the SINR of a multihoming user is not easily defined since SINR is not a simple additive quantity like throughput, thus to avoid confusion we do not include SINR values for multihoming scenarios in results or figures.} as a function of user type for the different network layout scenarios given the ER and TE resource schemes respectively. 

For the ER scheme, we find as expected that increasing the number of contracts (available for network switching) increases all three performance values of throughput, SINR and MOS. This result validates that exploiting the differences between multiple mobile networks through network switching can provide significant benefits. However surprisingly, we find that 3C-MH actually provides lower mean throughput and slightly lower MOS than 3C-NMH in all scenarios. In other words, multihoming does not provide on average an improved user experience compared to network switching with the same number of operator contracts.

Further investigation indicates that this is the result of inefficient radio resource allocation. Specifically, multihoming users typically have one strong signal and one weak signal, yet the networks allocate radio resources to both of these connections despite the fact that the strong connection often already provides high throughput to the user. Therefore the weak signal uses a large fraction of radio resources for a small and often unnecessary gain in throughput. If all users are in a similar situation then these weak connections result in an overall lower mean throughput. In other words, the problem is that each network lacks the information about the end user's multihoming nature to make intelligent radio resource allocation decisions.

\begin{figure}[!ht]
    \includegraphics[width=16.4cm]{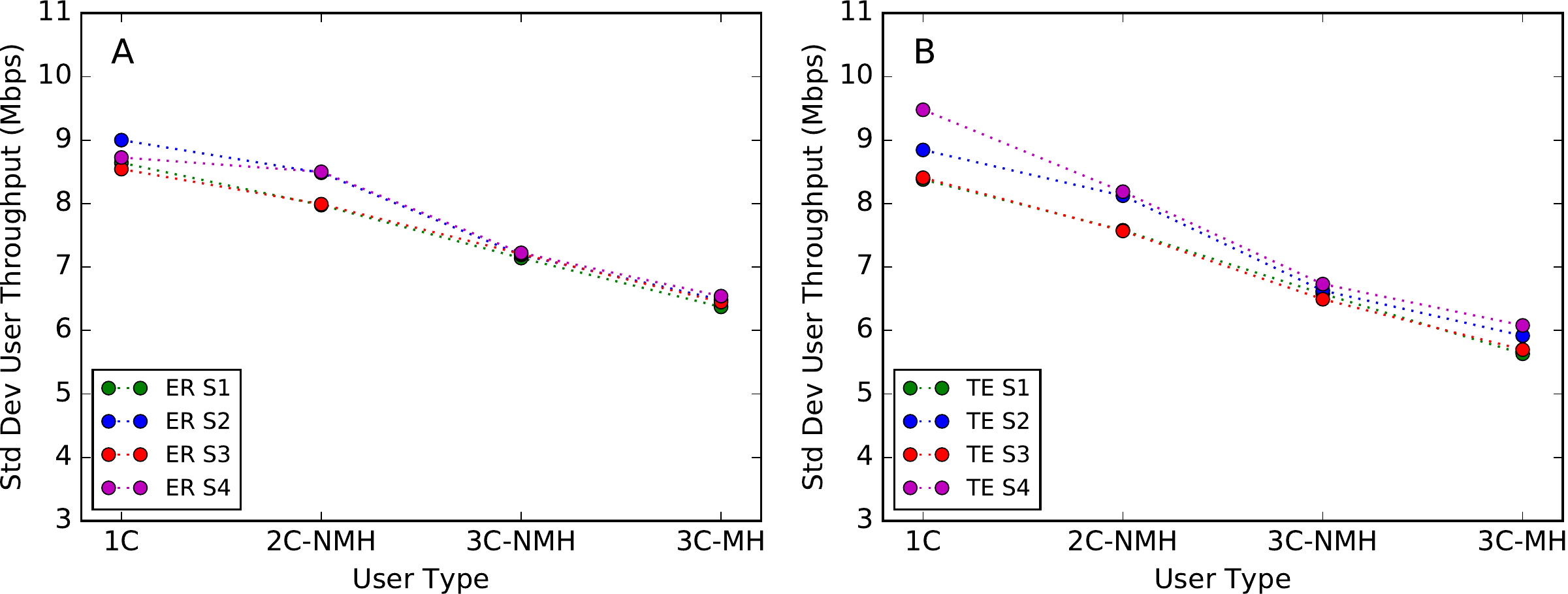}
    \caption{Standard deviation of throughput for different scenarios, user types (all users of a single type), and radio resource allocation schemes: A) ER allocation B) TE allocation.}
    \label{fig:std_er_te_all_scenarios} 
\end{figure}

However this lower throughput does not greatly effect MOS since the lower throughput is partly offset by a lower variation in throughput between users. Figure \ref{fig:std_er_te_all_scenarios} illustrates the standard deviation of throughput as a function of user type for different network layout scenarios.

In terms of individual scenarios, scenario 4 is a special case that shows unique dynamics because of the presence of an indoor only micro-operator (the green operator in figure \ref{fig:scenario3_4}B). For instance for 1C, the mean throughput of scenario 4 is the highest of all the scenarios but the MOS is the lowest of all the scenarios. This phenomenon is related to the high indoor throughput but very low (and often zero) outdoor throughput for users of the indoor only operator. Therefore this scenario numerically illustrates the potential problem for micro-operators without efficient network switching.

For the TE scheme, we find that although mean throughput decreases in some of the scenarios as the number of contracts increases, a more significant reduction in throughput variation drives MOS higher. Thus overall the pattern in terms of MOS is similar to the ER scheme. We also find a similar dynamic in terms of 3C-NMH compared to 3C-MH as in the ER scheme.

\begin{figure}[!ht]
    \includegraphics[width=16.4cm]{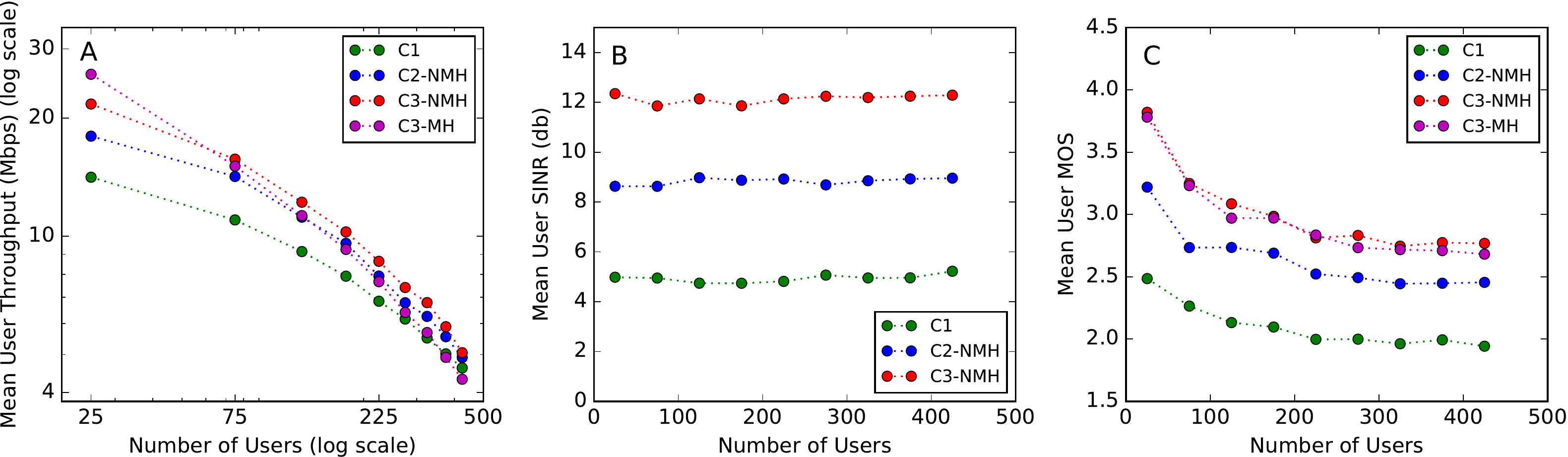}
    \caption{Performance values for different numbers of users and network switching and multihoming user types (all users of a single type) given ER radio resource allocation scheme and scenario 1: A) mean user throughput (log scale axes) B) mean user SINR C) mean user MOS.}
    \label{fig:num_users_scenario_1} 
\end{figure}

We also vary the number of users to determine if the previously mentioned inefficient radio resource allocation affects all user densities. Figure \ref{fig:num_users_scenario_1} illustrates the throughput, SINR, and MOS as a function of the number of users for different user types for network layout scenario 1 with ER scheme. Interestingly, we find that multihoming does provide better throughput for low density cases but that this advantage quickly disappears beyond about 75 users. Specifically, in the low density case inefficient radio resource allocation does not matter since there is very little competition for the resources. Thus selectively enabling end user multihoming depending on the congestion level of the networks might be a viable strategy. Though as also illustrated in Figure \ref{fig:num_users_scenario_1} the higher throughput of multihoming at low densities does not provide significantly higher MOS since the high throughput already available through network switching is adequate for the applications of the users.

\begin{figure}[!ht]
    \includegraphics[width=16.4cm]{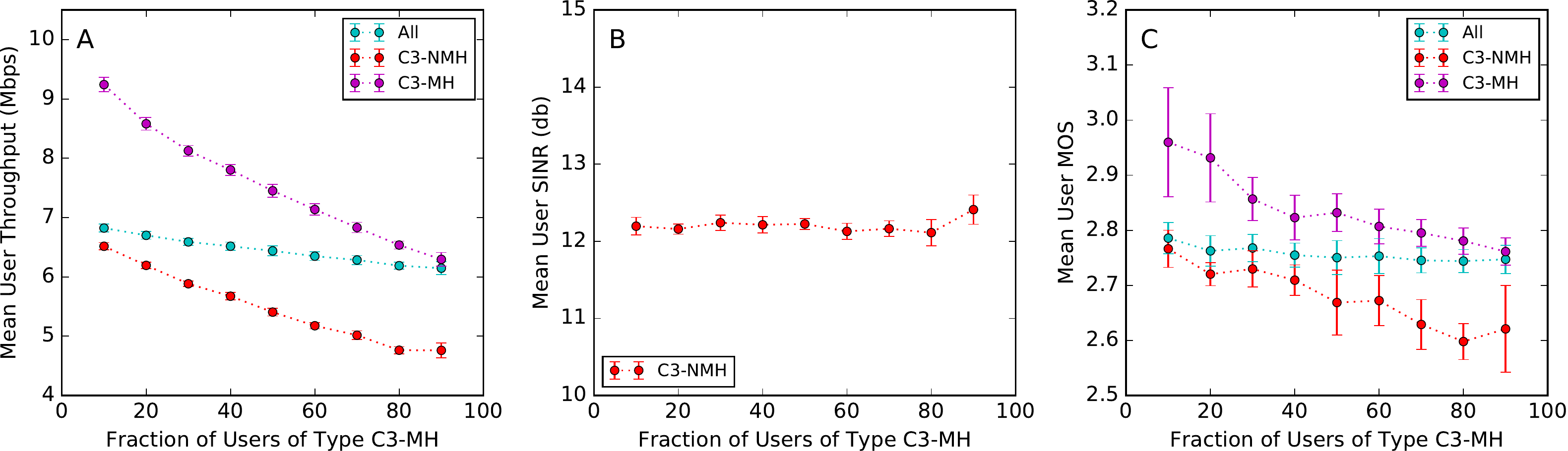}
    \caption{Performance values for 3C-MH and 3C-NMH users as a function of the fraction of users that are of the 3C-MH type given ER radio resource allocation scheme and scenario 1: A) mean user throughput B) mean user SINR C) mean user MOS.}
    \label{fig:frac_mh_scenario_1} 
\end{figure}

Additionally, we can vary the fraction of users of the different user types. Figure \ref{fig:frac_mh_scenario_1} illustrates the mean performance values for 3C-MH and 3C-NMH users as a function of the fraction of those users that are of the 3C-MH type (vs. 3C-NMH type) (also the mean of all users (collection of both types) is included for reference). We find that the advantage of 3C-MH users\footnote{In other words the distance between the 3C-MH and 3C-NMH curves in Figure \ref{fig:frac_mh_scenario_1}A and \ref{fig:frac_mh_scenario_1}C} (vs. 3C-NMH) is significant for both throughput and MOS and remains relatively constant as the fraction of 3C-MH users increases.

This observation is important for understanding potential adoption dynamics since late adopters will still have a significant incentive for adoption as their gain will be similar to early adopters. Comparatively, different dynamics are often at play in the adoption of new network technologies such as LTE in the sense that early adopters typically find non-congested networks while the network becomes congested as more users adopt the technology. In other words, the benefit of late adoption of LTE is less than early adoption.

Figure \ref{fig:frac_mh_scenario_1} also details that in general any amount of multihoming users decreases the performance of non-multihoming users. Again, this is the result of the inefficient radio resource allocation. Ref. \cite{dandachi2016} previously illustrated a similar problem in the case of multihoming and non-multihoming users over an LTE and a Wifi network (rather than several LTE networks). Though we also note that if the fraction of multihoming users is kept small then these users can share large gains with only minimal effect on non-multihoming users. Therefore, small special users groups, such as public safety officials, could use end user multihoming beneficially.

\subsection{Effects of network assumptions on results}\label{sec:network_assumptions}
As previously discussed, our network sub-model includes a basic radio access network but excludes backhaul, core network, and network protocol layers. Though the sub-model provides an appropriate level of detail for our analysis, such an approach naturally involves certain assumptions. We briefly discuss the potential impact of several of these network assumptions on our results. 

Firstly, since we do not model network protocols we simply assume that a network layer protocol can fully aggregate the capacity of several mobile connections (in other words simply sum the capacities). The most obvious candidate is MPTCP. In reality MPTCP cannot (even with optimal scheduling) fully aggregate capacities in situations with highly heterogeneous networks (in terms of throughput and delay) and limited buffer sizes \cite{choi2017,lim2017}. In other words, in situations especially like scenario 4 (indoor only operator) where users are often multihoming over a strong indoor and weak outdoor connection (or vice versa) the actual application layer throughput will be lower than our estimations. This phenomenon further reinforces the highly situational nature of benefits from end user multihoming. We look to include a model of MPTCP (or other multipath protocol) in future work.

Secondly, since we do not model backhaul or core network we do not consider potential congestion in these areas. Therefore, in scenarios where backhaul or core network are the limiting bottleneck our results, though still applicable, will be less important in terms of performance. Additionally, in such situations, the BS selection scheme would need to account for not only BS load but also backhaul or end-to-end load (similar to MPTCP schedulers when distributing data to sub-flows). Ref. \cite{ramazanali2016} summarizes several such association schemes. In any case, we believe in urban scenarios (like our simulations) the bottleneck is more likely to be the radio access network than backhaul or core network. Specifically, many urban areas often have, for example, expandable high capacity fiber-to-the-BS backhaul. 

Thirdly, we make certain radio access network assumptions such as constant transmission power of BSs. The constant transmission power assumption (along with other assumptions) implies no advanced inter-cell interference coordination. Therefore, we likely overestimate/underestimate the throughput of cell-center/cell-edge users compared to a simulation with, for example, soft or fractional frequency reuse \cite{yassin2017}. However, this does not significantly effect our results as all users move extensively over the simulation area and are equally affected. Additionally, even highly detailed LTE system level simulators such as the Vienna simulator \cite{rupp2016} do not yet support non-homogeneous power allocation.

\section{Discussion}\label{sec:discussion}
Interestingly, the results illustrate an important and (in hindsight) intuitive paradox for end user multihoming. Specially, a major benefit of end user multihoming, in that MNO support is not required, is also a limiting factor for the user benefit due to the need for such MNO support to perform efficient radio resource allocation for multihoming users. In the best case, operators could respond to the adoption of end user multihoming by working out a cooperative allocation system (especially given that without such support multihoming users may harm the network performance for all users). Such a cooperative scheme could work like the two level network-centric allocation\footnote{The first level scheduler would work at a flow level time scale (since inter-operator links would have significant delays) with approximate global information. Whereas the second level scheduler would work in each BS at millisecond time scale with information from the first level scheduler and additional, for example, fast fading information.} scheme described in Section II of \cite{dandachi2016}. However, if operators feel enabling multihoming might imperil their main business models then regulation may be needed to force such cooperation.

A potential indicator of MNO sentiment towards end user mechanisms and specifically network switching can be seen in the recent GSM Association rules regarding remote SIM provisioning affecting embedded-SIMs.\footnote{Embedded-SIMs are essentially reprogrammable SIMs that can hold several MNO profiles} The rules explicitly forbid the automated (without user intervention) switching between MNO profiles \cite{gsma2017}. This type of rule means that end user network switching likely becomes more technically challenging as embedded-SIM represented a relatively well developed potential enabler of network switching. Though we note that these provisioning rules are still subject to change.

Economically, the overall simulation results indicate that network switching improves allocative efficiency since the scarce radio resources and network capacity are better utilized. Though from a dynamic efficiency perspective (in other words in the long run), several authors \cite{hazlett2006,markendahl2011} have suggested that very low switching costs may decrease incentives for mobile network investment. For example, in a scenario with network switching where users have unlimited data plans, operators might be disincentivized from investing in areas already covered by a competitor (since it is more cost efficient to simply offload that traffic to the competitor). However, as mentioned, low switching costs also enable new small and medium sized operators to enter the market since they decreases entry barriers. As a result, the low costs incentivize investments in indoor and new network deployments, including IoT applications. 

Contrastingly, in a scenario with network switching where users have usage based data plans, operators might be incentivized to invest in areas with high data demands, in other words, they will compete for traffic. In such a scenario, regulators might ensure minimum coverage in areas with low population densities, while letting competition dominate in high population density areas. Overall, given the potential of network switching regulators should further investigate these incentive dynamics given the operators, pricing schemes, population densities, etc. in their local market.

\section{Conclusions}\label{sec:conclusions}
In this work we studied the benefits of end user network switching and end user multihoming in several diverse network scenarios through an agent based modeling approach. The model consists broadly of a detailed user behavior sub-model and high level mobile network sub-model and includes diverse performance metrics such as end user throughput, MOS, and SINR.

In term of results, the work indicates that end user network switching almost always improves mean user throughput and QoE as quantified by mean user MOS. Comparatively, end user multihoming only improves throughput and MOS in a smaller set of situations. Specifically, in complete multihoming adoption cases, multihoming does improve mean user throughput when user density is low (i.e. low congestion). Unfortunately in such low congestion situations, mean MOS is already high and thus additional throughput has only a marginal effect. 

In partial adoption cases, multihoming users have higher throughput and MOS (than non-multihoming users) and this advantage remains roughly constant even as the adoption fraction increases. Thus promoting the adoption of multihoming by late adopters might be less challenging than in other cases of network technology adoption, such as LTE. Though as mentioned the complete adoption of multihoming does not significantly improve user QoE (compared to the network switching case).

The main reason end user multihoming is not broadly beneficial is that, in end user multihoming, network operators are unaware that any given user is multihoming. This unawareness leads to networks making inefficient radio resource allocation decisions as the networks try to promote fairness between users. Since fairness is (rightly so) a major goal in most resource allocation algorithms, this effects essentially all resource allocation schemes.

Additionally, for network switching, the indoor operator scenario (scenario 4) quantifies the importance for indoor micro-operators of efficient network switching (so that their customers can easily switch to macro operators outdoors). Specifically, the mean user MOS of these micro-operator customers roughly doubles given network switching.

Overall, this work supports the general adoption of end user network switching for increasing allocative efficiency and the adoption of end user multihoming in limited situations. In addition, we hope our results spur more research into areas such as algorithms for dynamically enabling multihoming in certain situations, mechanisms for cooperative (inter-network) resource allocation of multihoming users, and the benefits of further lowering MNO switching costs.

\section*{Acknowledgment}
The work of Benjamin Finley was supported by the EMERGENT Project (http://emergent.comnet.aalto.fi/). The work of Arturo Basaure was supported by the Fondecyt project \# 11170100. We also want to thank Kalevi Kilkki, Heikki H{\"a}mm{\"a}inen, Antti Oulasvirta, Jukka Manner, Pasi Lassila, Alexandr Vesselkov, and Jaume Benseny for providing comments and feedback on this work.

%\section*{References}
\bibliography{ms}
\section*{Appendix}
\subsection{User model distributions}

\begin{table}[!ht]
\centering
\caption{Video application session resolution probabilities.}
\label{tab:video_resolutions}
\begin{tabular}{|l|l|}
\hline
\textbf{Resolution} & \textbf{Probability} \\ \hline
360 & 10\% \\ \hline
480 & 10\% \\ \hline
720 & 10\% \\ \hline
1080 & 20\% \\ \hline
2000 & 30\% \\ \hline
4000 & 20\% \\ \hline
\end{tabular}
\end{table}

\begin{sidewaystable*}
\small
\centering
\begin{threeparttable}
\caption{Summary of distributions and functions in user behavior and QoE model.}
\label{tab:user_behavior_model}
\begin{tabular}{|l|l|l|l|l|l|l|}
\hline
& & \textbf{Web Browsing} & \textbf{Video} & \textbf{Facebook} & \textbf{Messaging} & \textbf{Maps/Navigation} \\ \hline
\multirow{2}{*}{Activity Arrival}& Model & \multicolumn{5}{c|}{Poisson Process (min)}\\ \cline{2-7} 
& Assump & \multicolumn{5}{c|}{Following successful activity: $\lambda_{s}=N(0.10,0.03)$, Following failed activity: $\lambda_{f}=N(0.05,0.03)$} \\ \hline
\multirow{2}{*}{Additional Sess in Activity} & Model & \multicolumn{5}{c|}{Bernoulli distribution (after session)}\\ \cline{2-7} 
& Assump & \multicolumn{5}{c|}{Following complete session: $p_{s}=N(0.50,0.05)$, Following abandon session: $p_{f}=N(0.20,0.05)$}\\ \hline
\multirow{2}{*}{Session Arrival} & Model & \multicolumn{5}{c|}{Generalized Pareto (sec)}\\ \cline{2-7} 
& Assump & \multicolumn{5}{c|}{$K=1.31$, $\sigma=11.48$, $\theta=0.0$}\\ \hline
\multirow{2}{*}{Session Duration\tnote{a}}& Model &Weibull\tnote{b} (sec) & Weibull\tnote{b} (sec)&Weibull\tnote{b} (sec)&Log Normal (sec) & Log Normal (sec)\\ \cline{2-7} 
& Assump &$\alpha=0.46$, $\beta=34.40$ & $\alpha=0.39$, $\beta=88.21$ & $\alpha=0.37$, $\beta=18.15$ & $\mu=2.49$, $\theta=1.70$ & $\mu=2.91$, $\theta=1.94$\\ \hline
\multirow{2}{*}{Session Abandon Prob}& Model & \multicolumn{5}{c|}{Bernoulli distribution (every sec)}\\ \cline{2-7} 
& Assump &\multicolumn{5}{c|}{${p_A} = \left\{{\begin{array}{*{20}{c}}{0.083 - 0.033*({\rm{MOS) }}}&{{\rm{if}}\ {t_s} > {T_a}\ \rm{and}\ \rm{MOS} \le 2.5}\\
0&{{\rm{if}}\ {t_s} \le {T_a}\ \rm{or}\ \rm{MOS} > 2.5}\end{array}} \right\}$}\\ \hline
Session Initial Startup & Assump & \multicolumn{5}{c|}{$T_{a}=5  \rm{sec}$}\\ \hline
\multirow{2}{*}{User Session Selection Prob\tnote{c}} & Model & Normal & Normal & Normal & Normal & Normal\\ \cline{2-7} 
& Assump & $\mu=0.20$, $\sigma=0.10$ & $\mu=0.20$, $\sigma=0.10$ & $\mu=0.30$, $\sigma=0.10$ & $\mu=0.30$, $\sigma=0.10$ & $\mu=0.05$, $\sigma=0.10$\\ \hline
\multirow{2}{*}{Goodput To MOS} & Model & \multicolumn{5}{c|}{B-Splines}\\ \cline{2-7} 
& Assump & \multicolumn{5}{c|}{Linear basis function with knots derived from reported results in \cite{casas2016a} and \cite{casas2016b}}\\ \hline
\end{tabular}
    \begin{tablenotes}
      \small
      \item[a] Session duration distributions are estimated empirically via powerlaw package \cite{alstott2014} on the active smartphone dataset described in \cite{finley2016b}. In each case the distribution is selected based on Akaike weights.
      \item[b] Weibull distributions are parametrization as in equations 1 and 2 in \cite{weisstein2016}.
      \item[c] This probability of each new session being a certain type is itself drawn from the stated probability distributions with given parameters. Thus different users have different probabilities of selecting each session type whenever they start a new session. This helps introduce user diversity in the distribution of user's time between session types.
    \end{tablenotes}
\end{threeparttable}
\end{sidewaystable*}

\end{document}